\documentclass{PoS}
\usepackage{lineno}
\usepackage{bbm}
\usepackage{amsmath}
\usepackage{verbatim}
\title{Dirac Eigenvalue spectrum of $N_f$=2+1 QCD towards the chiral limit using HISQ fermions}

\ShortTitle{Dirac Eigenvalue spectrum of $N_f$=2+1 QCD towards the chiral limit using HISQ fermions}

\author{Heng-Tong Ding$^1$, Olaf Kaczmarek$^{1,2}$, Frithjof Karsch$^{2}$, {Sheng-Tai Li}$^{3,1}$, Swagato Mukherjee$^4$, Akio Tomiya$^5$, \speaker{Yu Zhang}$^1$ \thanks{This work was supported by the National Natural Science Foundation of China under grant numbers 11775096, 11535012, 11947237, and the Deutsche Forschungsgemeinschaft (DFG, German Research Foundation) - project number 315477589 - TRR 211, the RIKEN Special Postdoctoral Researcher program, the U.S. Department of Energy, Office of Science, Office of Nuclear Physics through the Contract No. DE-SC0012704
and the U.S. Department of Energy, Office of Science, Office of Nuclear Physics and Office of Advanced Scientific Computing Research within the framework of Scientific Discovery through Advance Computing (SciDAC) award Computing the Properties of Matter with Leadership Computing Resources. The numerical simulations have been performed on the GPU cluster in the Nuclear Science Computing Center at Central China Normal University (NSC$^3$), Wuhan, China. We thank the HotQCD collaboration for sharing the gauge configurations.} 
\\
	\llap{$^1$}	Key Laboratory of Quark \& Lepton Physics (MOE) and Institute of Particle Physics,  \\
	Central China Normal University, Wuhan 430079, China. \\
	\llap{$^2$}	Fakult\"at f\"ur Physik, Universit\"at Bielefeld, D-33615 Bielefeld, Germany\\
	\llap{$^3$} Institute of Modern Physics,Chinese Academy of Sciences, Lanzhou 730000, China\\
    \llap{$^4$} Physics Department, Brookhaven National Laboratory, Upton, NY 11973, USA\\
      \llap{$^5$} RIKEN/BNL Research center, Brookhaven National Laboratory, Upton, NY, 11973, USA\\
E-mail:
\email{hengtong.ding@mail.ccnu.edu.cn},
\email{okacz@physik.uni-bielefeld.de},
\email{karsch@physik.uni-bielefeld.de},
\email{stli@impcas.ac.cn},
\email{swagato@bnl.gov},
\email{akio.tomiya@riken.jp},
\email{yuzhang@mails.ccnu.edu.cn}
}

\abstract{ We utilize the eigenvalue filtering technique combined with the stochastic estimate of the mode number to determine the eigenvalue spectrum. Simulations of (2 + 1)-flavor QCD are performed using the Highly Improved Staggered Quarks (HISQ/tree) action on $N_{\tau}$ = 8  lattices with aspect ratios $N_{\sigma}/N_{\tau}$ ranging from 5 to 7. The strange quark mass is fixed to its physical value $m_{s}^{\rm phy}$, and the light quark masses $m_{l}$ are varied from $m_{s}^{\rm phy}/40$ to $m_{s}^{\rm phy}/160$ which correspond to pion mass $m_{\pi}$ ranging from 110 MeV to 55 MeV in the continuum limit.
We compute the chiral condensate and $\chi_{\pi} - \chi_{\delta}$ through the eigenvalue spectrum obtained from the the eigenvalue filtering method.
We compare these results with those obtained from  a direct calculation of the observables which involves inversions of the fermion matrix using the stochastic "noise vector" method. We find that these approaches yield consistent results. Furthermore, we also investigate the quark mass and temperature dependences of the Dirac eigenvalue density at zero eigenvalues to gain more insights about the $U_A(1)$ symmetry breaking in QCD.} %with the purpose to gain more insight about the $U_A(1)$ symmetry breaking in QCD. 

\FullConference{37th International Symposium on Lattice Field Theory - Lattice2019\\
		16-22 June 2019\\
		Wuhan, China}

\begin{document}

\section{Introduction}
\vspace{-0.1cm}
The Lagrangian of Quantum chromodynamics (QCD) with $N_f$ flavors posseses a $U(N_f)_L\times U(N_f)_R \equiv SU(N_f)_L\times SU(N_f)_R \times U_A(1)\times U_V(1) $ symmetry in the limit of vanishing quark masses. The $SU(N_f)_L\times SU(N_f)_R$ chiral symmetry is spontaneously broken in the vacuum and gets restored above the chiral phase transition temperature \cite{Ding:2019prx}. However, the axial $U_A(1)$ symmetry is always broken due to the axial anomaly. The degree of its breaking is expected to play a significant role in determining the order of the chiral phase transition \cite{Pisarski:1983ms}. Thus a thorough understanding of the fate of the $U_A(1)$ symmetry at high temperature is very important. %Thus it is important to understand the fate of $U_A(1)$ symmetry at high temperature. 

Several attempts have been made to address this problem using lattice QCD simulations in $N_f$=2+1 \cite{Buchoff:2013nra,Dick:2015twa,Bazavov:2019www} and $N_f$=2 QCD \cite{Brandt:2016daq,Suzuki:2017ifu} over the last few years. The general conclusion is that at pseudo critical temperature the $U_A(1)$ symmetry remains broken. In the present work we explore the microscopic mechanisms responsible for $U_A(1)$ symmetry breaking in the high temperature by studying the low-lying eigenvalue spectrum of staggered fermions on the (2 + 1)-flavor HISQ configurations with small quark masses towards the chiral limit and large volumes.
   
The eigenvalue spectrum $\rho(\lambda,m) = T/V \langle\sum_{k}\delta(\lambda_{k}(m) - \lambda)\rangle$ is a probe of the spontaneous chiral symmetry breaking through the Banks-Casher relation \cite{Banks:1979yr}:
\begin{small}
\begin{equation}
 \lim_{\lambda \to 0}\lim_{m \to 0}\lim_{V \to\infty}\pi\rho(\lambda,m) = \lim_{m \to 0}\lim_{V \to \infty} \langle\bar{\psi}\psi\rangle 
 \end{equation}
 \end{small}
 Thus, $\rho(0,0) \neq 0$ means a nonzero value of chiral condensate in the chirally broken phase and $\rho(0,0)=0$ leads to a vanishing chiral condensate in the chirally symmetric phase. The chiral condensate can be expressed in terms of the eigenvalue spectrum of the Dirac operator as
\begin{small}
\begin{equation}
\langle \bar{\psi}\psi \rangle \stackrel{V \rightarrow \infty}{\longrightarrow}  \frac{N_{f}}{4}\int_{0}^{\infty} {\rm d} \lambda \frac{2m\rho(\lambda,m)}{\lambda^2 + m^2} .
\label{eq.test21}
\end{equation}
\end{small}
It indicates that the low-lying eigenvalue spectrum contribute the most for the breaking and restoration of the chiral symmetry. Whether $\rho(\lambda,m)$ opens a gap in the low-lying mode indicating the $U_A(1)$ symmetry restoration is another important question. As an observable to characterize the $U_A(1)$ symmetry breaking , the $U_A(1)$ susceptibility $\chi_{\pi} - \chi_{\delta}$ is defined by the difference of the integrated correlation functions between pion and delta meson and it can be given in terms of the eigenvalue spectrum of the Dirac operator as 
 \begin{small}
 \begin{equation}
 \chi_{\pi} - \chi_{\delta} 
 =  \frac{N_{f}}{4} \int {\rm d}^4x[\langle i\pi^\dagger(x)i\pi^-(0)\rangle - \langle \delta^\dagger(x)\delta^-(0)\rangle]
 \stackrel{V \rightarrow \infty}{\longrightarrow}  \frac{N_{f}}{4}\int_{0}^{\infty}{\rm d} \lambda \frac{4m^2\rho(\lambda,m)}{(\lambda^2 + m^2)^2} .
 \label{eq.test22}
 \end{equation}
 \end{small}

\vspace{-0.2cm}
In this work we utilize the Chebyshev filtering technique combined with the stochastic estimate method \cite{NapoliPS13, deForcrand:2017cja, Cossu:2016yzp, Fodor:2016hke}  to calculate the eigenvalue spectrum and the mode number of the staggered Dirac operator in lattice gauge theories. Instead of using the commonly used method which is Kalkreuter-Simma(KS) Ritz algorithm \cite{Kalkreuter:1995mm} to directly compute the individual low-lying eigenvalues, the idea is to estimate the number of eigenvalues located in any given interval. One great advantage is that we can get the whole eigenvalue spectrum in a computationally inexpensive way\footnote{For example, the Chebyshev filtering method with polynomial order p=24000 costs 3 minutes for one configuration with one noise vector on $56^3\times8$ lattices with $m_{\pi}$ = 55 MeV and costs same for different quark masses, while calculating 192 low-lying eigenvalues the KS algorithm costs around 3.5 hours for one same configuration on a single V100 GPU and the computing time increases as the quark mass decreases.}. In the next section we will discuss the Chebyshev filtering method in detail. 
\section{Chebyshev filtering}
\vspace{-0.1cm}
Stochastic counting of eigenvalues of a hermitian matrix $A$ in a given interval [s,t] within [-1,1] can be represented as
\begin{small}
\begin{equation}
n[s,t] \simeq \frac{1}{N_r} \sum_{r=1}^{N_r}\xi_{r}^{\dagger}h(A)\xi_r ,
\label{eq.test1}
\end{equation}
 \end{small}
 %\vspace{-0.1cm}
where $\xi_{r}$ is a Gaussian random noise vector, ${N_r}$ is the number of random vectors and $h(A)$ is a step function which equals to 1 only in the interval [s,t] and 0 elsewhere. In practice, the function $h(A)$ is approximated by the Chebyshev polynomial
\vspace{-0.1cm}
\begin{small}
\begin{equation}
h(A) \simeq \sum_{j=0}^{p}g_j^{p}\gamma_{j}T_{j}(A).
\label{eq.test2}
\end{equation}
 \end{small}
 \vspace{-0.1cm}
The coefficients $g_j^{p}$ and $\gamma_{j}$ are known numbers once the interval [s,t] is given, and $p$ is the order of Chebyshev polynomials. As the expansion of $h(A)$ has harmful oscillations near the boundaries $g_j^{p}$ is introduced \cite{NapoliPS13} here to suppress this behavior. $T_{j}(A)$ is the Chebyshev polynomial of operator $A$ and it can be constructed by the following recursion relation
\begin{small}
\begin{equation}
T_0(A) = 1,\quad T_1(A) = A,\quad T_j(A) = 2AT_{j-1}(A) - T_{j-2}(A) \quad(j\geq 2 ).
\label{eq.test3}
\end{equation}
\end{small}
 The above deviation is based on the assumption that all the eigenvalues of $A$ are restricted in the range of [-1,1]. In order to apply the eigenvalue filtering method to calculate the Dirac spectrum, we therefore define
\begin{small}
\begin{equation} 
   A = \frac{D^{\dagger}D - \frac{(\lambda^{D^{\dagger}D}_{{\rm max}} + \lambda^{D^{\dagger}D}_{{\rm min}})}{2} \mathbbm{1} }{\frac{(\lambda^{D^{\dagger}D}_{{\rm max}} + \lambda^{D^{ \dagger} D}_{{\rm min}})}{2} \mathbbm{1}},
  \label{eq.test4}
\end{equation}
\end{small}
  such that the eigenvalues of $A$ are all distributed in [-1,1]. 
  
  Substituting the expression of $h(A)$ (Eq.~(\ref{eq.test2})) into the stochastic estimator (Eq.~(\ref{eq.test1})) and after averaging over the gauge fields, we can obtain the mode number $\bar{n}[s,t]$
\begin{small}
  \begin{equation}
  \bar{n}[s,t] \approx \frac{1}{N_r} \sum_{r=1}^{N_r}  \sum_{j=0}^p g^{p}_j \gamma _{j} \langle{ \xi_r^\dagger T_{j}(A) \xi_r }\rangle.
    \label{eq.test5}
  \end{equation}
  \end{small}
  Once we get the mode number, the spectral density can be easily constructed as
    \begin{small}
   \begin{equation}
  \rho(\lambda, m) = \frac{1}{2V}\frac{\bar n[s,t]}{\delta},
    \label{eq.test6}
   \end{equation}
   \end{small}
   where the factor of 2 in the denominator is due to pairs of negative and positive eigenvalues, and $\delta$ is the binsize.  $\lambda$ stands for the eigenvalue of $A$ 
    and it is related to the positive eigenvalues of Dirac operator $iD$ as follows,
  \begin{small}
   \begin{equation}
    \lambda^{\vert iD \vert} = \sqrt{\lambda^{D^{\dagger}D}} = \left[\left(\lambda^{D^{\dagger}D}_{\rm max} - \lambda^{D^{ \dagger} D}_{\rm min}\right)s/2 +  \left(\lambda^{D^{\dagger}D}_{\rm max} + \lambda^{D^{ \dagger} D}_{\rm min}\right)/2 \right]^{1/2},
    \end{equation}
    \end{small}
    \vspace{-0.2cm}
    \begin{small}
    \begin{equation}
     \lambda^{\vert iD \vert}+\delta= \sqrt{\lambda^{D^{\dagger}D}}+\delta = \left[ \left(\lambda^{D^{\dagger}D}_{{\rm max}} - \lambda^{D^{ \dagger} D}_{{\rm min}}\right)t/2  +  \left(\lambda^{D^{\dagger}D}_{\rm max} + \lambda^{D^{ \dagger} D}_{\rm min}\right)/2 \right]^{1/2},
  \end{equation}
  \end{small}
   where $\lambda^{D^{\dagger}D}_{{\rm min}}$ is set to 0 and  $\lambda^{D^{\dagger}D}_{{\rm max}}$ is estimated by the power method.  
\section{Lattice setup}
\vspace{-0.1cm} 
Our simulations of (2 + 1)-flavor QCD are performed using the Highly Improved Staggered Quarks (HISQ/tree) action on $N_{\tau}$ = 8 lattices with aspect ratios $N_{\sigma}/N_{\tau}$ ranging from 5 to 7 in order to keep $m_{\pi}L$ fixed to reduce the finite volume effects. In the current simulation the strange quark mass is fixed to its physical value $m_{s}^{\rm phy}$, and the light quark masses $m_{l}$ are varied from $m_{s}^{\rm phy}/40$ to $m_{s}^{\rm phy}/160$ which correspond to pion mass $m_{\pi}$ ranging from 110 MeV to 55 MeV in the continuum limit. We use about 1000 configurations at each temperature and each value of the quark mass, where each configuration is separated by 20 time units for $m_{s}^{\rm  phy}/40$ and 10 times units for other quark masses after skipping the first thermalized 1000 time units of each stream for decorrelation.
\section{Mode number and reproduction of chiral observables from the eigenvalue spectrum}
\begin{figure}[h!]	
\vspace{-0.4cm}		
	\begin{center}
		\includegraphics[width=0.42\textwidth]{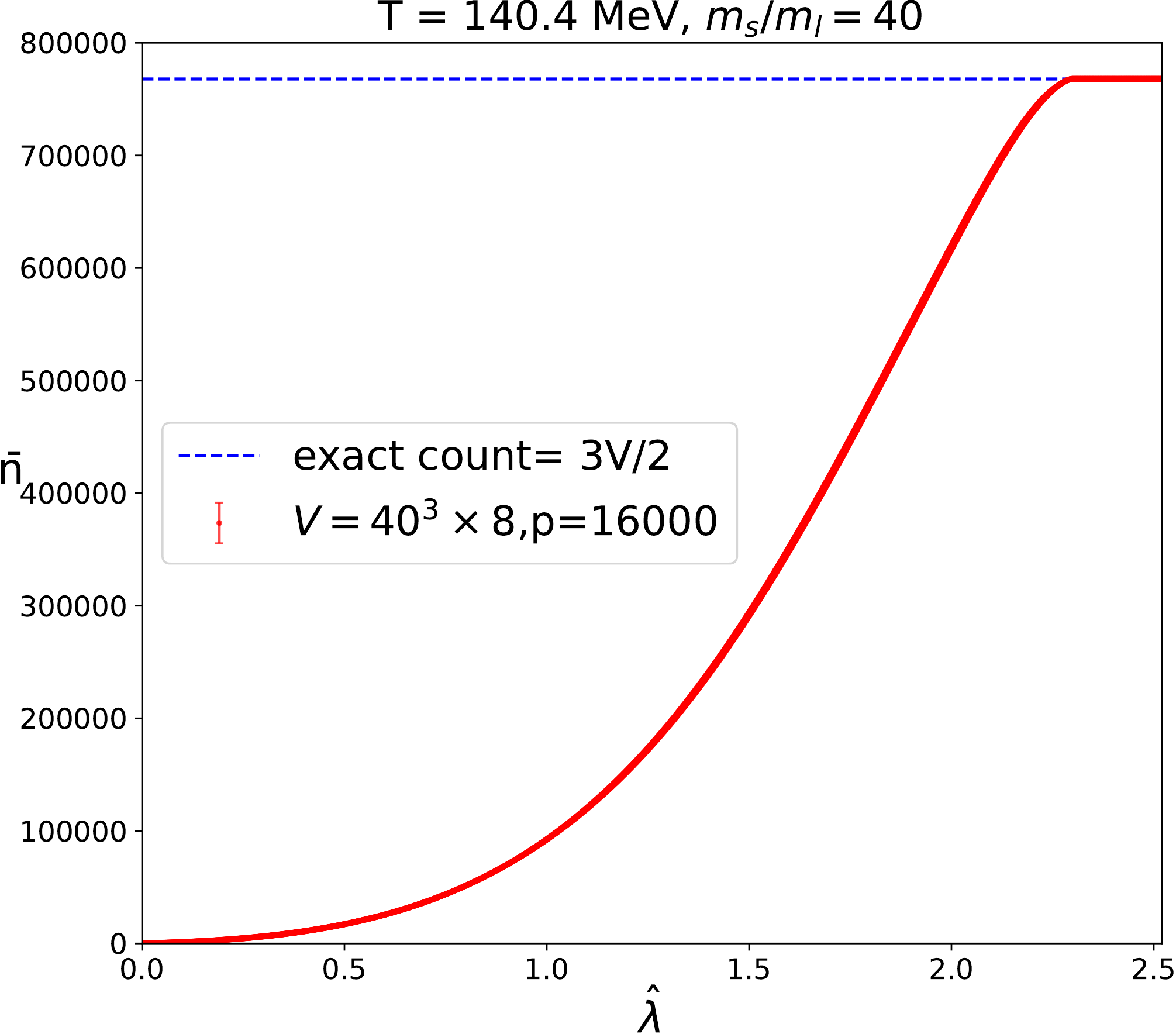}~
		\includegraphics[width=0.405\textwidth]{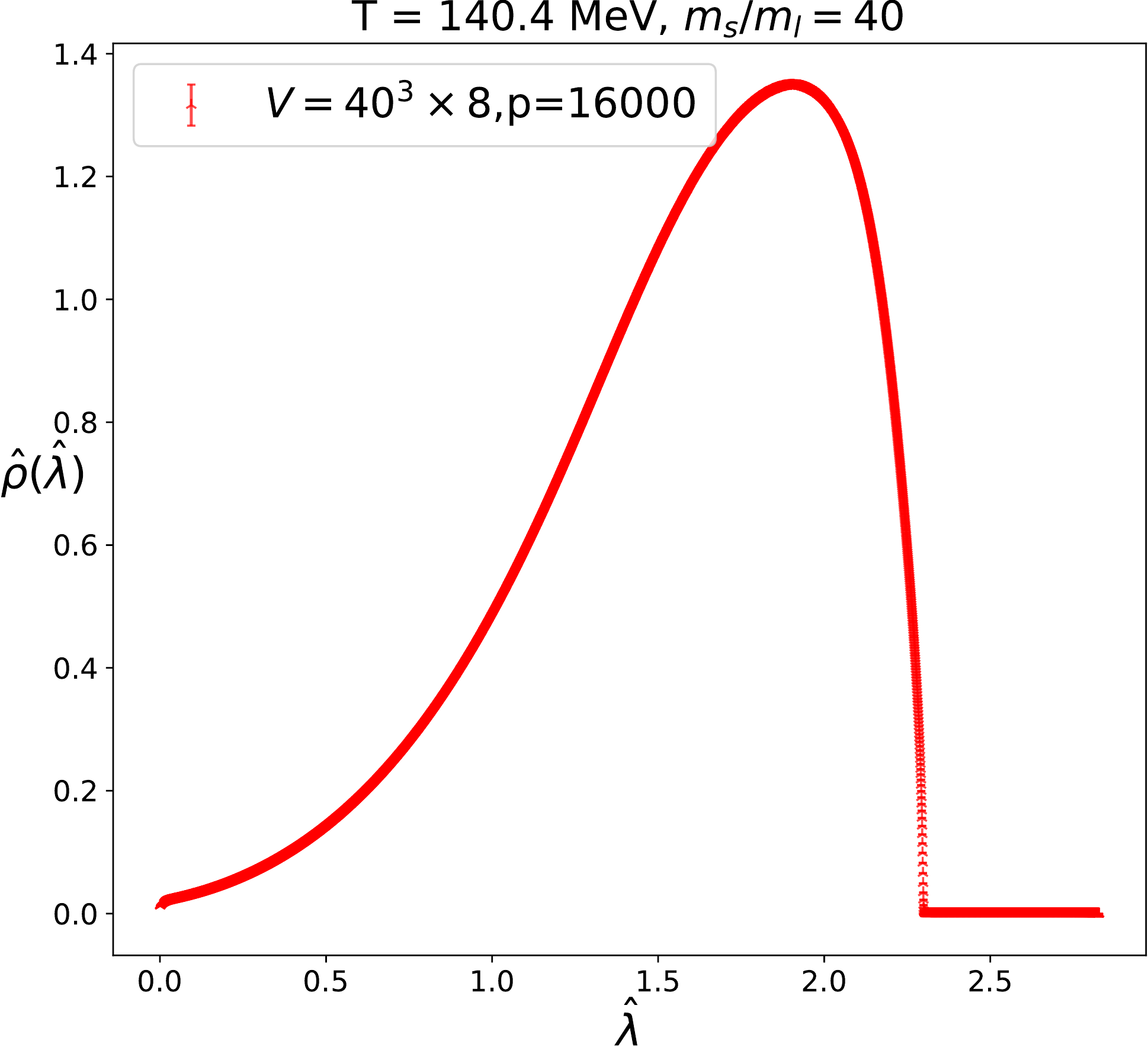}
	\end{center}
	\vspace{-0.4cm}
	\caption{ Left: The mode number distribution of the staggered Dirac operator from the Chebyshev expansion with the Chebyshev polynomial order p=16000 on $40^3\times8$ lattices and it perfectly matches to the exact number of eigenvalues which denoted as the blue line in the large $\hat{\lambda}\equiv\lambda a$ region. Right: The eigenvalue spectrum $\hat{\rho}(\hat{\lambda}) \equiv \rho(\lambda a)a^3$ obtained from the mode number as shown in the left plot.}	 
	\label{fig.1}	
\end{figure}
To demonstrate the Chebyshev filtering approach, we show in the left panel of Fig.~\ref{fig.1} the mode number distribution of the staggered Dirac operator $\vert iD \vert$ obtained using the Chebyshev polynomial expansion method (c.f.Eq.~(\ref{eq.test5})). Chebyshev polynomials up to an order of 16000 and 20 random noise vectors are used on $40^3\times8$ lattices at T=140.4 MeV and the quark mass is set as $m_s^{\rm phy}/40$. It converges to the correct total number of eigenvalues $\frac{3V}{2}$ in the large $\hat{\lambda}\equiv\lambda a$ region. Here 3 is the $N_c$ factor, and $V$ is the full volume of the lattice, as we only count the positive eigenvalues we also need to divide by 2. Using the mode number we are be able to determine the eigenvalue spectrum according to Eq.~(\ref{eq.test6}). The eigenvalue spectrum is computed by binning of eigenvalues with a binsize of 0.0005 as shown in the right panel of Fig.~\ref{fig.1}.

 Once the eigenvalue spectrum is obtained, we proceed to compare the chiral condensate $\langle\bar{\psi}\psi\rangle$ and $\chi_{\pi} - \chi_{\delta}$ computed from the eigenvalue spectrum with those obtained from a direct calculation of the observables which involves inversions of the fermion matrix using the stochastic noise vector approach. In the stochastic noise vector approach, these chiral observables are estimated by using noise vectors $\eta_k$, $k$ = 1, .., $n$, via
 \begin{small}
\begin{equation} 
\langle \bar{\psi}\psi \rangle = \frac{N_{f}}{4}\frac{1}{V} \langle  {{\rm Tr} M^{-1}} \rangle \approx \frac{N_{f}}{4}\frac{1}{V} \langle \frac{1}{N_r}\sum_{r=1}^{N_r} \xi_{r}^{\dagger} M^{-1} \xi_{r} \rangle,
 \label{eq.test9}
  \end{equation}
   \end{small}
   \begin{small}
    \begin{equation} 
     \chi_{\pi} - \chi_{\delta} = \frac{N_{f}}{4}\frac{1}{mV} \langle{{\rm Tr} M^{-1}} \rangle + \frac{N_{f}}{4}\frac{1}{V} \langle {\rm Tr} M^{-2}  \rangle 
     \approx \frac{N_{f}}{4}\frac{1}{mV} \langle \frac{1}{N_r}\sum_{r=1}^{N_r} \xi_{r}^{\dagger} M^{-1} \xi_{r} \rangle + \frac{N_{f}}{4}\frac{1}{V} \langle  \frac{1}{N_r}\sum_{r=1}^{N_r} \xi_{r}^{\dagger} M^{-2} \xi_{r} \rangle.
     \label{eq.test10}
      \end{equation}
      \end{small}
\vspace{-0.3cm}
\begin{figure}[h!]			
	\begin{center}
		\includegraphics[width=0.421\textwidth]{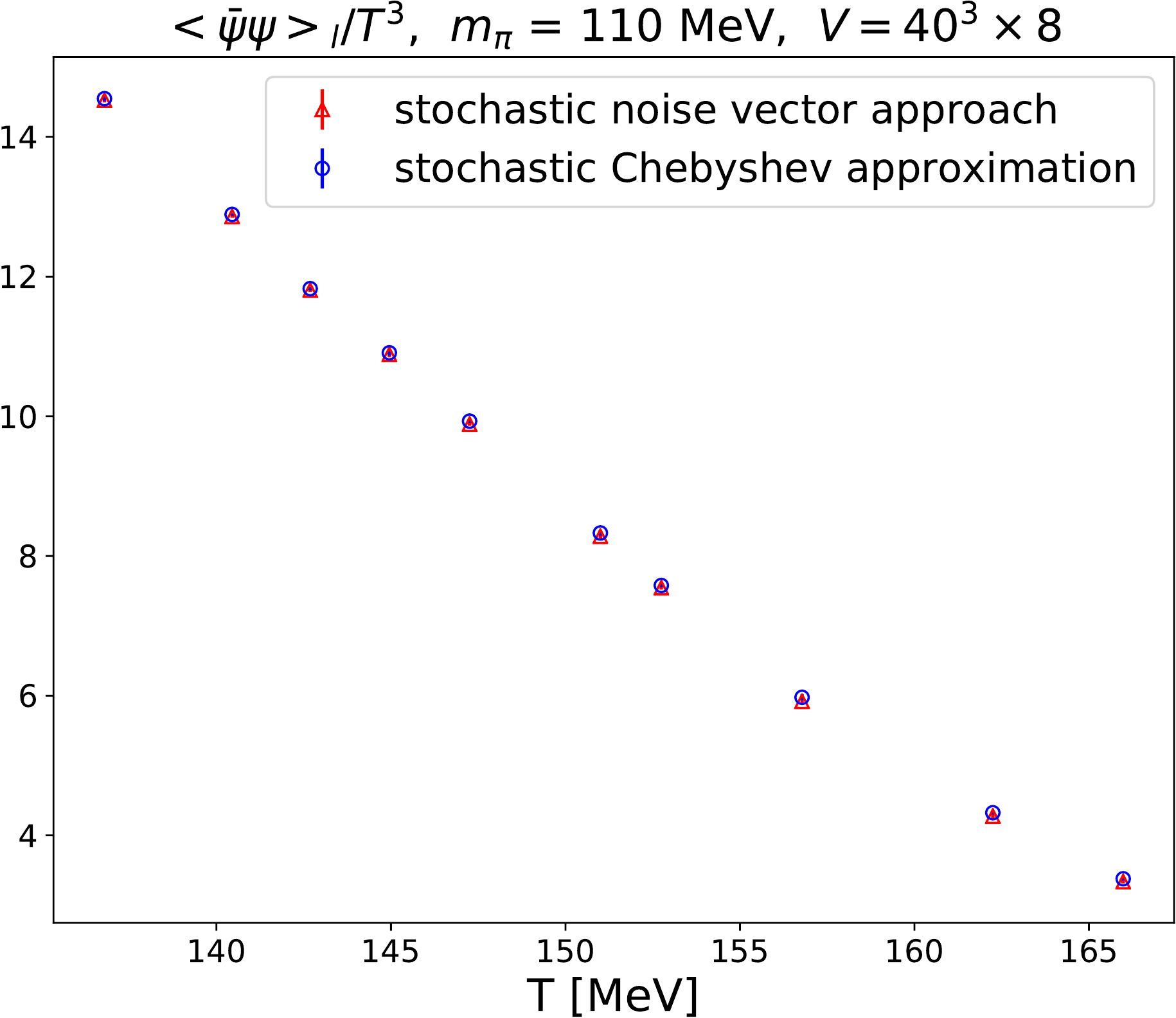}~
		\includegraphics[width=0.43\textwidth]{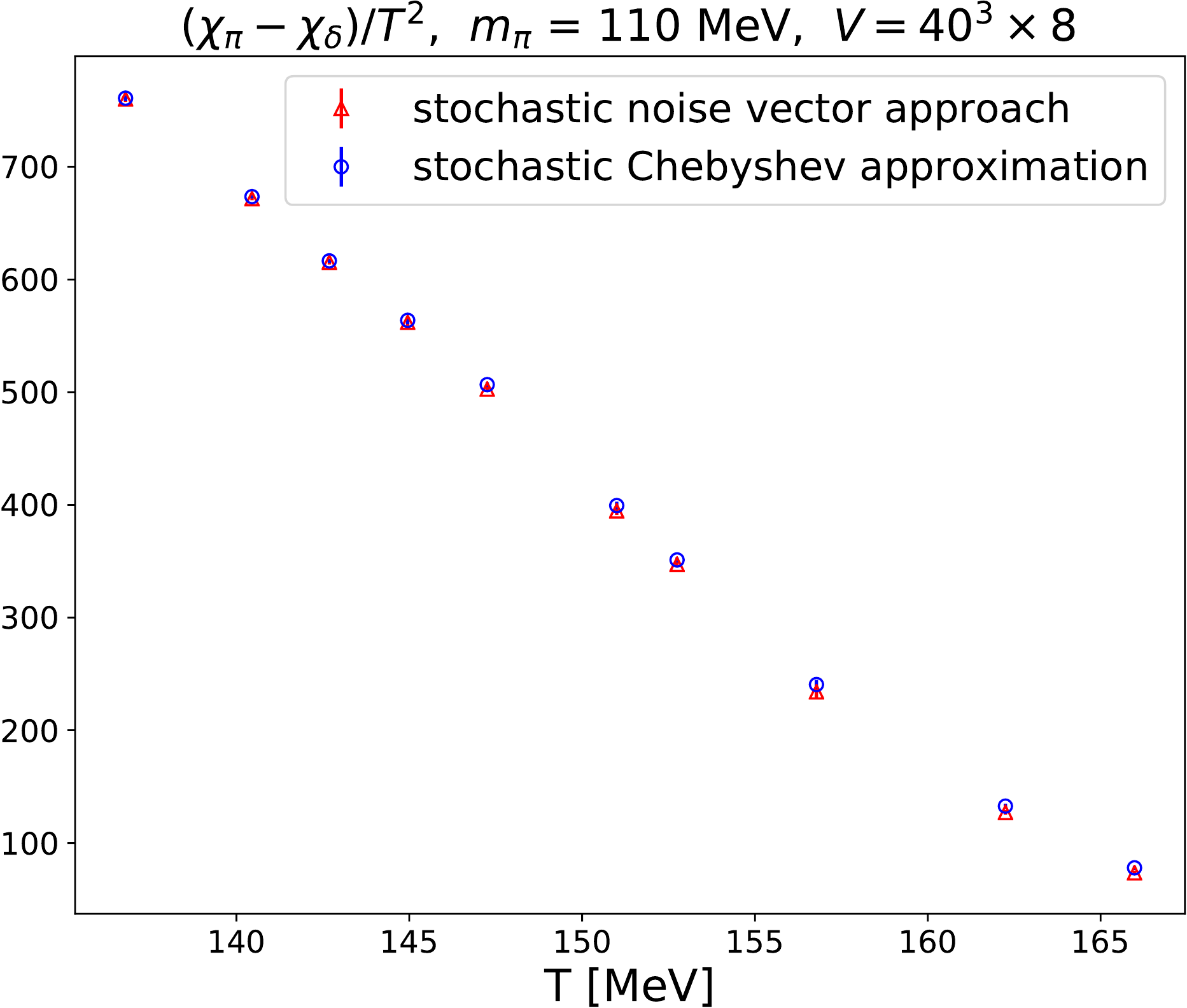}~
	\end{center}
\vspace{-0.2cm}
\caption{The temperature dependence of $\langle\bar{\psi}\psi\rangle_l/T^3$(left) and $(\chi_{\pi} - \chi_{\delta})/T^2$ (right) obtained from the eigenvalue spectrum (blue) versus that are calculated from the exact inversion of the Dirac operator using stochastic noise vectors (red) at $m_{\pi}$ = 110 MeV on $40^{3}\times8$ lattices.}
	\label{fig.2}	
\end{figure}

Fig.~\ref{fig.2} shows the quantitative comparison of $\langle\bar{\psi}\psi\rangle_l/T^3$ and $(\chi_{\pi} - \chi_{\delta})/T^2$ at different temperatures obtained on $40^{3}\times8$ lattices with $m_{\pi}$ = 110 MeV. These values are obtained on the same configurations with the same number of random vectors. As seen from the left plot of Fig.~\ref{fig.2} $\langle\bar{\psi}\psi\rangle_l/T^3$ obtained from these approaches agree very well with each other. This is also the case for $(\chi_{\pi} - \chi_{\delta})/T^2$ as shown in the right plot of Fig.~\ref{fig.2}. These consistency from both approaches is not clearly observed on earlier studies \cite{Buchoff:2013nra} using Domain Wall fermions.
\section{The eigenvalue spectrum and the fate of $U_A(1)$ symmetry}
\begin{figure}[h!]			
	\begin{center}
		\includegraphics[width=0.433\textwidth]{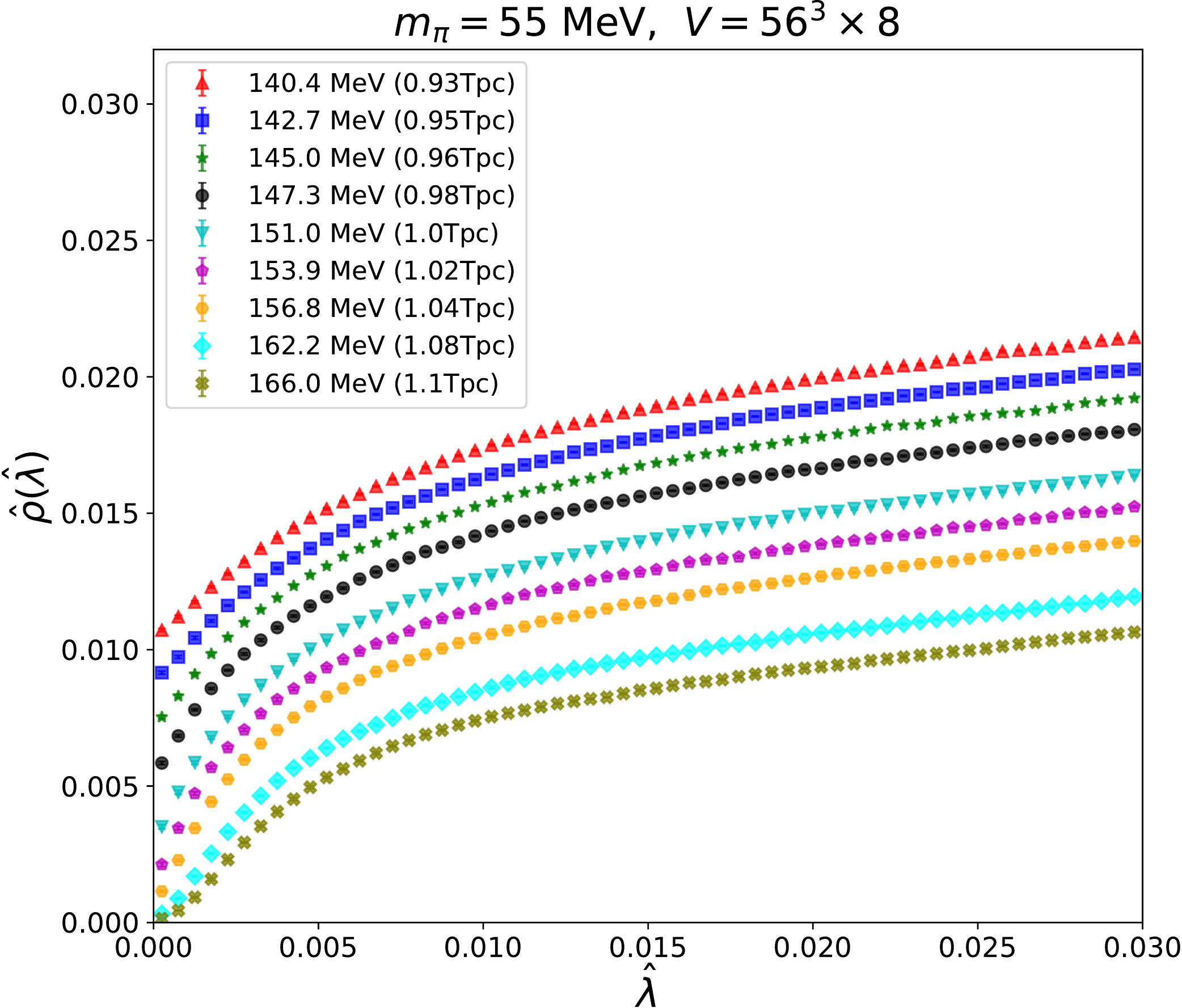}~
		\includegraphics[width=0.42\textwidth]{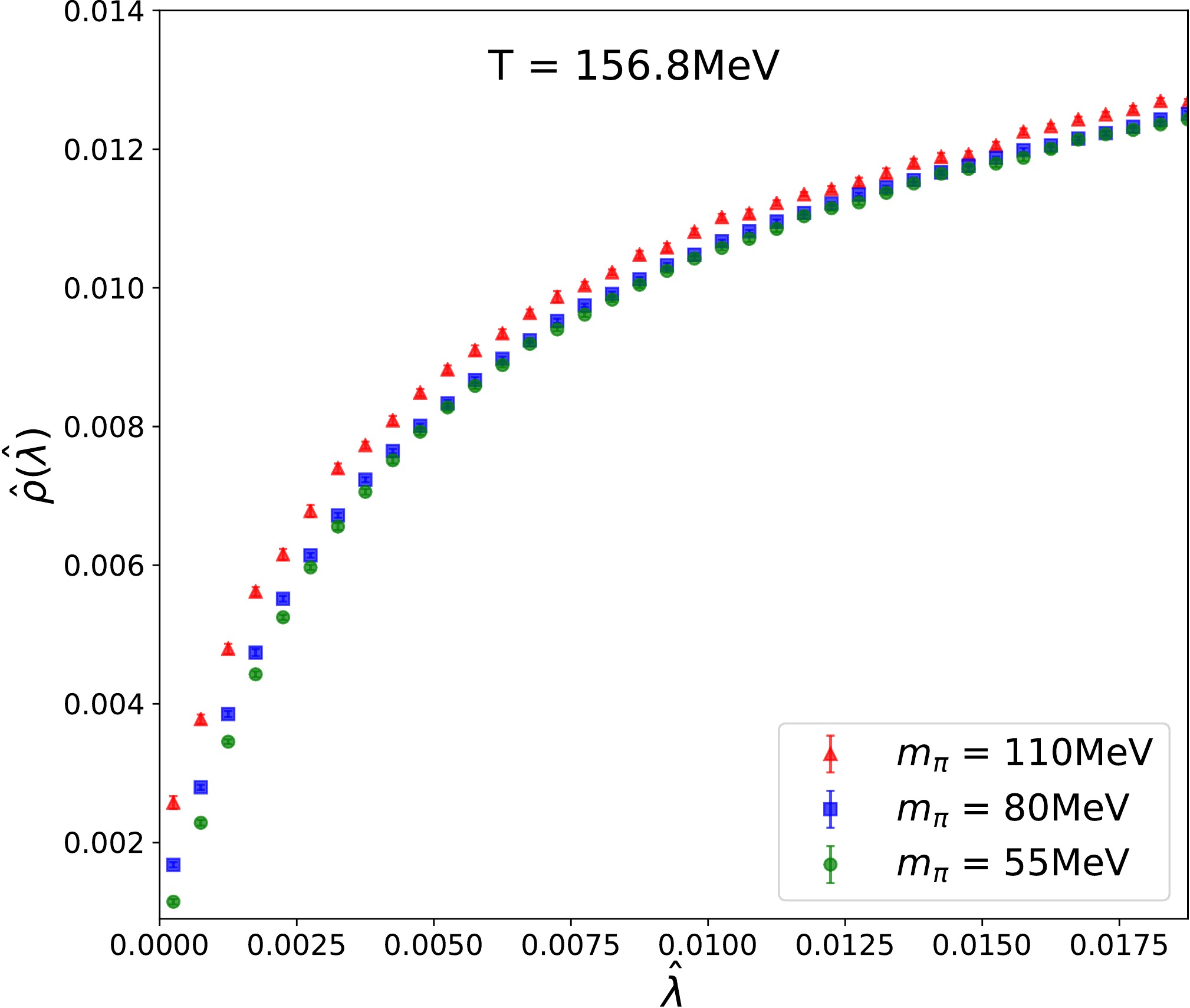}
	\end{center}
	\vspace{-0.3cm}
	\caption{The left figure shows the temperature dependence of the low-lying eigenvalue spectrum at quark mass $m_{s}^{\rm phy}/160$. The right figure shows the quark mass dependence at a fixed temperature. }	 
	\label{fig.3}	
\end{figure}

\vspace{-0.1cm}
We calculate the low-lying eigenvalue spectrum of staggered Dirac operator with various light quark masses at different temperatures. The temperature dependences of $\hat\rho(\hat\lambda, \hat m)$ at our smallest light quark mass $m_{s}^{\rm phy}/160$ ($m_{\pi}$ = 55 MeV) is shown in the left panel of Fig.~\ref{fig.3}. We find that as the temperature increases the near zero modes are suppressed, and finally approaches to zero. This is expected from the chiral symmetry restoration above the pseudo critical temperature $T_{pc}$. Another observation is that there is no clear evidence of a gap around zero up to our current highest temperature 166 MeV. The mass dependence of the eigenvalue spectrum is shown in the right panel of Fig.~\ref{fig.3}. We found a relatively larger quark mass dependence at small eigenvalue part and a smaller quark mass dependence at large eigenvalue part.

\begin{figure}[h!]			
	\begin{center}
		\includegraphics[width=0.42\textwidth]{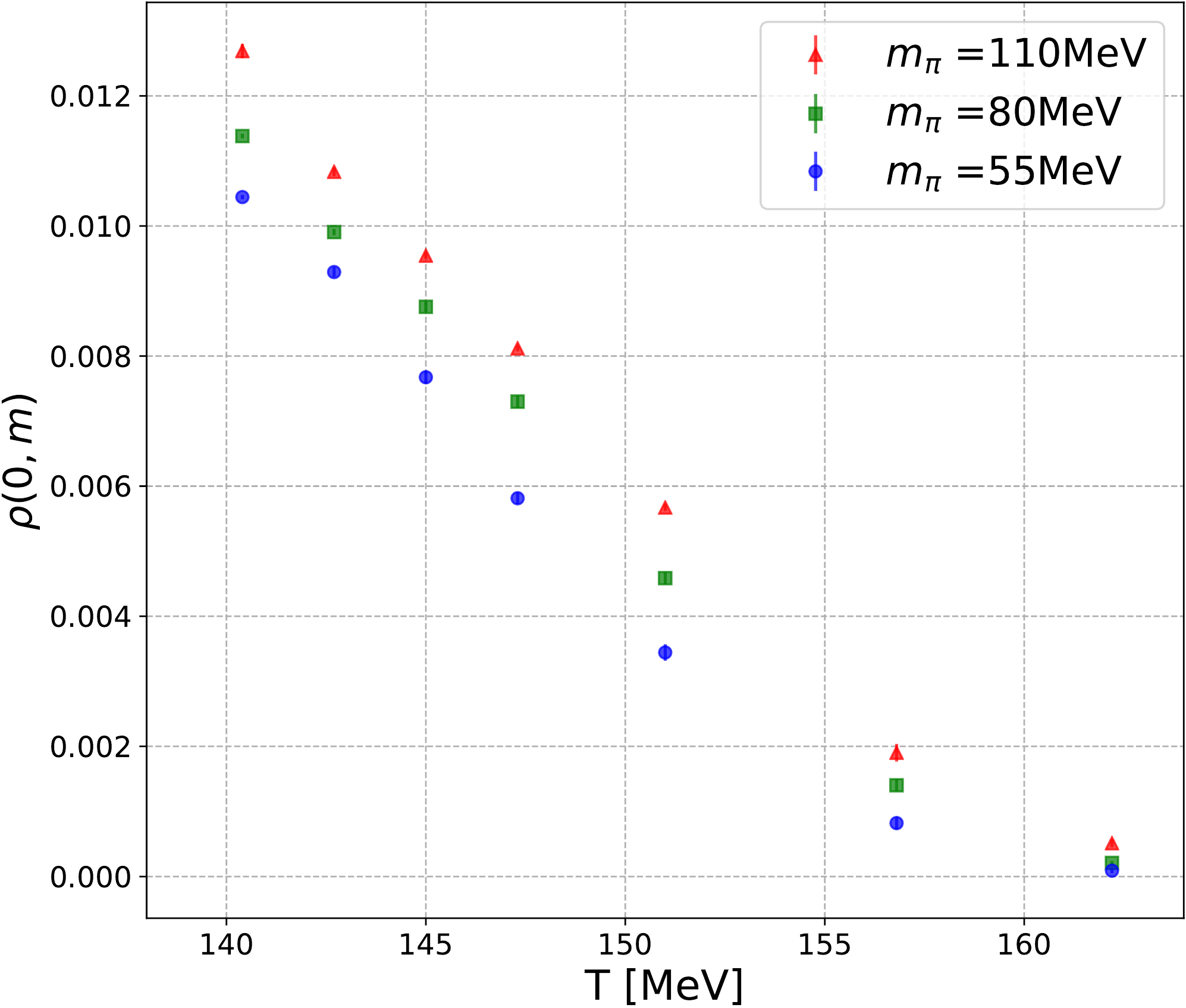}~
			\includegraphics[width=0.433\textwidth]{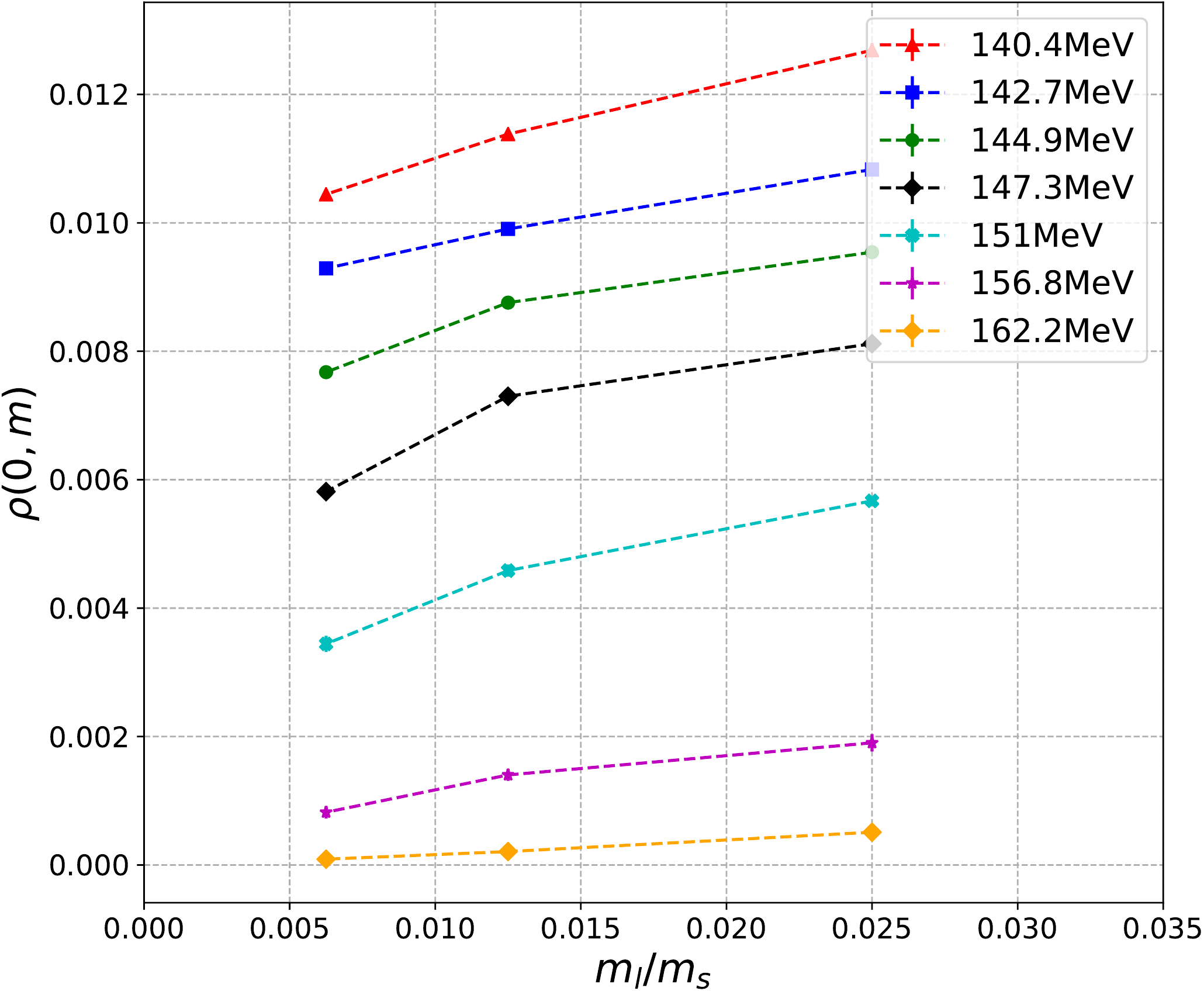}
	\end{center}
	\vspace{-0.3cm}
	\caption{The Dirac eigenvalue spectrum at $\lambda = 0$ versus temperature (left) and quark mass (right).}	 
	\label{fig.4}	
\end{figure}

Next we try to examine the behavior of the $\rho(\lambda,m)$ at $\lambda$ = 0. It has been shown in \cite{Ohno:2012br, Bazavov:2012qja} that above the critical temperature if $\rho(0,m) = cm$ ,  $\langle\bar{\psi}\psi\rangle$ vanishes in the chiral limit while $\chi_{\pi} - \chi_{\delta} = \chi_{{\rm disc}} \neq 0$. To estimate $\rho(0,m)$ we fit the eigenvalue spectrum to a cubic polynomial function in $\lambda$
 \begin{equation} 
 \rho(\lambda,m) = c_0 + c_1 \lambda + c_2 \lambda^2 + c_3\lambda^3. 
  \label{eq.test11}
   \end{equation}
   So $c_0$ is $\rho(0,m)$. In the left plot of Fig.~\ref{fig.4} the behavior of the $\rho(0,m)$ at different temperature is shown. It can be seen that $\rho(0,m)$ monotonically decreases as the temperature increases and it seems to vanish at around 162 MeV as $SU(2)_L\times SU(2)_R$ tends to be restored towards the chiral limit.
   The right plot of Fig.~\ref{fig.4} shows the quark mass dependence of $\rho(0,m)$ at different temperatures. It shows that $\rho(0,m)$ seems to have a linear dependence on the mass at 162.2 MeV which indicates that the $U_A(1)$ symmetry remains broken.

\section{Conclusions}
\vspace{-0.1cm}
The eigenvalue filtering technique utilized in this work is proven to be very effective to obtain the eigenvalue spectrum of the staggered Dirac operator. Using the eigenvalue spectrum we successfully reproduced the chiral condensate and $\chi_{\pi} -\chi_{\delta}$, which are obtained from the stochastic noise vector approach.

We have investigated the temperature and quark mass dependences of the eigenvalue spectrum and found that the eigenvalue spectrum at zero eigenvalues $\rho(0,m)$ seems to go to zero at 162.2 MeV towards the chiral limit suggesting the restoration of the $SU(2)_L\times SU(2)_R$ chiral symmetry. We didn't find any evidence for a gap near $\lambda=0$ in the eigenvalue spectrum up to 166 MeV at our smallest quark mass. A linear dependence of $\rho(0,m)$  on the quark mass found at $\sim$ 162 MeV may be connected to the breaking of the $U_A(1)$ symmetry.

%%%%%%%%%%%%%%%%%%%%%%%%%%%%%%%%%%%%%%%%%%%%%%%%%%%%%%

\bibliographystyle{JHEP}
\bibliography{lattice2019_yuzhang.bib}

\end{document}